\documentclass[%
print,
 amsmath,amssymb,
 aps,
]{revtex4}

\usepackage{times}%此指令是令文章字体选用Palation.
\usepackage{graphicx}% Include figure files
\usepackage{dcolumn}% Align table columns on decimal point
\usepackage{bm}% bold math
\usepackage{dsfont}
\usepackage{mathrsfs}
\usepackage[landscape,papersize={297.1mm,210mm},left=1.9cm,right=1.6cm,top=2.7cm,bottom=2.8cm]{geometry}% 此指令是规定A4纸的页边距
\usepackage[                   %dvipdfm, pdflatex,pdftex这里决定运行文件的方式不同
            pdfstartview=FitH,
            colorlinks, %注释掉此项则交叉引用为彩色边框(将colorlinks和pdfborder同时注释掉)
            pdfborder=001,   %注释掉此项则交叉引用为彩色边框
            linkcolor=blue,
            anchorcolor=blue,
            citecolor=blue,
            urlcolor=blue
            ]{hyperref}
\usepackage{amsthm}
\makeatletter

\newcommand{\Rmnum}[1]{\expandafter\@slowromancap\romannumeral #1@}
\newcommand{\tabincell}[2]{\begin{tabular}{@{}#1@{}}#2\end{tabular}}
\makeatother

\begin{document}

\title{Teleporting an unknown qutrit state via a 2-qudit entangled channel}

\author{Xiao-Xu Li$^{1}$}
\author{Feng-Li Yan$^{1}$}
\email{flyan@hebtu.edu.cn}
\author{Ting Gao$^{2}$}
\email{gaoting@hebtu.edu.cn}

\affiliation {$^1$ College of Physics, Hebei Key Laboratory of Photophysics Research and Application, Hebei Normal University, Shijiazhuang 050024, China\\
$^2$ School of Mathematical Sciences, Hebei Normal University, Shijiazhuang 050024, China}

\begin{abstract}
We propose a quantum teleportation scheme for transmitting a single qutrit state by adopting a 2-qudit entangled state as the quantum channel. The measurement basis for Alice has been carefully and systematically constructed, which is essential for the successful implementation of the teleportation protocol. Based on Alice's measurement outcomes, we design the corresponding collective unitary transformations to be performed by Bob on an auxiliary qubit and information particle. After implementing the collective unitary transformation, Bob performs a von Neumann measurement on the auxiliary qubit. The single qutrit state is then teleported to the distant receiver Bob with a finite success probability. We obtain the achievable success probabilities of the proposed teleportation scheme for different quantum channels. The obtained results not only enrich the theory of quantum teleportation over high-dimensional entangled channels but also provide a novel and feasible approach to implementing qutrit teleportation.~\\


\end{abstract}

\maketitle

\section{Introduction}

Quantum mechanics has fundamentally reshaped our understanding of the encoding, processing, and transmission of information. By exploiting uniquely quantum features such as entanglement, quantum systems can accomplish information processing tasks that are impossible in classical systems. These distinctive quantum properties form the foundation of quantum information science and continue to drive extensive research in areas such as quantum cryptography \cite{mimaxue1,mimaxue2,mimaxue3,mimaxue4}, quantum communication \cite{tongxin2,tongxin3,tongxin1,tongxin4}, and quantum algorithms \cite{jisuan1,jisuan2,jisuan3,jisuan4}. Among various quantum resources, entanglement plays a central role in quantum information processing and serves as a key physical resource for many quantum information protocols, including quantum teleportation \cite{Bennett,teleportation1,teleportation2}, quantum dense coding \cite{densecoding1,densecoding2,densecoding3}, and quantum cryptographic communication protocols \cite{mimaxue1,tongxin3,keydistribution2,keydistribution3,securedirectcommunication,
secretsharing1,secretsharing2}. Among these applications, quantum teleportation is of particular importance and is widely regarded as one of the most remarkable applications of quantum physics in information theory.

Quantum teleportation was first proposed by Bennett et al. \cite{Bennett} in 1993. To jointly implement quantum teleportation, the sender Alice and the receiver Bob in different locations pre-share a pair of maximally entangled two-particle quantum states to establish the required quantum channel. Alice performs a joint measurement on the particle with the unknown quantum state to be teleported and one particle in the maximally entangled state. Bob executes a unitary operation on his particle according to Alice's measurement results. Thereby, the unknown quantum state can be perfectly reconstructed on Bob's particle. The concept sparked widespread research interest and spurred landmark studies that extended the standard protocol to probabilistic teleportation \cite{gailv1,gailv2,gailv3,gailv4,gailv5}, controlled quantum teleportation \cite{controlled1,controlled2,controlled3,controlled4} and high-dimensional quantum teleportation \cite{HD1,HD2}, with remarkable theoretical \cite{teleportation1,Bennett,gailv1,gailv2,gailv3,gailv4,gailv5,controlled1,controlled2,controlled3,controlled4,HD1} and experimental \cite{HD2,shiyan1,shiyan2,shiyan3} progress. By introducing a controller into Bennett's protocol, Karlsson et al. \cite{controlled1} proposed the controlled quantum teleportation protocol in 1998. In 2000, Li et al. \cite{gailv1} introduced an auxiliary particle into their scheme, achieving probabilistic quantum teleportation with a fidelity of 1. Roa et al. \cite{gailv2} presented an information-lossless quantum teleportation protocol in 2015, where no information of the unknown quantum state is lost even if teleportation fails, i.e., this teleportation protocol allows repeated attempts without cloning the original state. In recent years, with the growing attention to high-dimensional quantum systems, high-dimensional quantum teleportation has gradually emerged as an important research direction in quantum information science.

Recently, increasing attention has been devoted to high-dimensional quantum teleportation due to its advantages in enhancing the information capacity and security of quantum communication. In particular, quantum teleportation based on qutrit systems \cite{Qutritsystems1,Qutritsystems2} has been extensively studied. However, in many existing teleportation schemes, the pre-shared quantum channels are usually required to be maximally entangled states, which to some extent limits the choice of quantum channels. To address this issue, Chen et al. \cite{chen} proposed a perfect quantum teleportation scheme in 2022. Their scheme employs two partially entangled 2-qutrit pure states with equal maximum Schmidt coefficients as the quantum channel, which enables the perfect teleportation of an unknown qutrit state. Unlike conventional schemes, their work indicates that the quantum channel is not restricted to maximally entangled states but can be extended to a continuous region within the entanglement invariant space. This result provides new insights for further exploring quantum teleportation schemes in high-dimensional quantum systems.

Motivated by the work of Chen et al., we further investigate quantum teleportation in high-dimensional quantum systems and extend their study to higher-dimensional settings. In this paper, we propose a quantum teleportation scheme for teleporting an unknown qutrit state by employing a 2-qudit entangled pure state as the quantum channel. Compared with low-dimensional entangled systems, quantum channels constructed from higher-dimensional entangled states may possess richer entanglement structures and more flexible state correlations. onsequently, their application in quantum teleportation is expected to offer significant advantages in terms of transmission capacity \cite{capacity}, quantum logic gate execution efficiency \cite{logicgate}, communication security and robustness against noise \cite{communicationsecurityandnoiseresistance1,communicationsecurityandnoiseresistance2}.

This paper is organized as follows. In Sec. \ref{Q2}, we present a quantum teleportation framework employing a 2-qudit entangled state as the quantum channel and construct the corresponding projective measurements performed by the sender Alice. In Sec. \ref{Q3}, we derive the appropriate unitary operations that enable the receiver Bob to reconstruct the unknown quantum state, thereby achieving perfect quantum teleportation. In Sec. \ref{Q4}, we provide a discussion and concluding remarks. Our work not only enriches the theoretical study of high-dimensional quantum teleportation, but also offers a new theoretical pathway for teleportation based on two-qudit entangled channels.

\section{Teleportation of an unknown qutrit state via a 2-qudit entangled channel}\label{Q2}

Suppose that the sender Alice wants to teleport an arbitrary unknown qutrit state of information particle 1,
\begin{equation}
|\psi\rangle_{1}=\alpha|0\rangle_{1}+\beta|1\rangle_{1}+\gamma|2\rangle_{1}
\end{equation}
to the receiver Bob, where Alice and Bob are in the different places, $\alpha, \beta$ and $\gamma$ are complex numbers, and satisfy the normalization condition $|\alpha|^{2}+|\beta|^{2}+|\gamma|^{2}=1$. Additionally, we also assume that Alice and Bob share a 2-qudit entangled state of particles 2 and 3,
\begin{equation}
|\Phi\rangle_{23}=a_{0}|00\rangle_{23}+a_{1}|11\rangle_{23}+a_{2}|22\rangle_{23}+a_{3}|33\rangle_{23}
\end{equation}
as a quantum channel. Here the Schmidt coefficients $\{a_{i}|i=0,1,2,3\}$ are non-negative real number that satisfies $0\leq a_{0}\leq a_{1}\leq a_{2}\leq a_{3}$ and $\sum_{i=0}^{3}a_{i}^{2}=1$, and the particles 1 and 2 belong to the sender Alice, particle 3 belongs to the receiver Bob, each of particles 2 and 3 is the 4-level particle. We expect that by using $|\Phi\rangle_{23}$ as the quantum channel, one can achieve perfect transmission of the single-qutrit state $|\psi\rangle_{1}$.

The quantum state $|\Phi\rangle_{23}$ can be prepared from quantum state $|00\rangle_{23}$ of the particles 2, 3 by using two unitary operations
\begin{equation}
\begin{aligned}
U_{1} =\left(\begin{array}{cccc}
        a_{0} & a_{1} & a_{2} & a_{3}\\
        a_{1} & -a_{0} & -a_{3} & a_{2}\\
        a_{2} & a_{3} & -a_{0} & -a_{1}\\
        a_{3} & -a_{2} & a_{1} & -a_{0}
\end{array}
\right),
\end{aligned}
\end{equation}
and
\begin{equation}
\begin{aligned}
U_{2}=|0\rangle_{2}\langle0|\otimes I_{3}+|1\rangle_{2}\langle1|\otimes V_{3}^{\dagger}+|2\rangle_{2}\langle2|\otimes W_{3}+|3\rangle_{2}\langle3|\otimes V_{3},
\end{aligned}
\end{equation}
where $I_{3}$ is the identity operator of particle 3 and $V_{3}=|0\rangle\langle1|+|1\rangle\langle2|+|2\rangle\langle3|+|3\rangle\langle0|$, $W_{3}=|0\rangle\langle1|+|1\rangle\langle2|+|2\rangle\langle0|+|3\rangle\langle3|$. It is easy to check that
\begin{equation}\nonumber
U_{1}|0\rangle_2=a_{0}|0\rangle_{2}+a_{1}|1\rangle_{2}+a_{2}|2\rangle_{2}+a_{3}|3\rangle_{2}
\end{equation}
and
\begin{equation}\nonumber
U_2U_{1}|00\rangle_{23}=a_{0}|00\rangle_{23}+a_{1}|11\rangle_{23}+a_{2}|22\rangle_{23}+a_{3}|33\rangle_{23}=|\Phi\rangle_{23}.
\end{equation}
So after the implement of unitary operators $U_1$ and $U_2$, the quantum state $|00\rangle_{23}$ is transformed into $|\Phi\rangle_{23}$. By sending the particle 3 to the receiver Bob, the quantum channel $|\Phi\rangle_{23}$ required for this quantum teleportation is perfectly constructed.

In Ref. \cite{chen}, the authors shown that in order to perfectly transfer the unknown quantum state $\alpha'|0\rangle+\beta'|1\rangle$ (the fidelity of quantum teleportation is 1, and the success probability is 1) the following two conditions must be met: (i) Projective measurement is performed by Alice; (ii) The collapsed state of particle 3 can be locally transformed into the form of $\alpha'|\widetilde{0}\rangle+\beta'|\widetilde{1}\rangle$, where $|\widetilde{0}\rangle$ and $|\widetilde{1}\rangle$ are two orthogonal states independent of the normalization coefficients $\alpha'$ and $\beta'$. For the teleportation of a single qutrit state $\alpha|0\rangle+\beta|1\rangle+\gamma|2\rangle$, with condition (i) remaining unchanged, we rewrite condition (ii) as the collapsed state of particle 3 can be locally transformed into the form of $\alpha|\widetilde{0}\rangle+\beta|\widetilde{1}\rangle+\gamma|\widetilde{2}\rangle$, where $|\widetilde{0}\rangle$, $|\widetilde{1}\rangle$ and $|\widetilde{2}\rangle$ are three orthogonal states independent of $\alpha$, $\beta$ and $\gamma$.

The total tripartite state of the system can be written as
\begin{align}
\nonumber
|\Psi\rangle_{123}&=|\psi\rangle_{1}|\Phi\rangle_{23}\\
\nonumber
&=(\alpha|0\rangle+\beta|1\rangle+\gamma|2\rangle)_{1}(a_{0}|00\rangle+a_{1}|11\rangle+a_{2}|22\rangle+a_{3}|33\rangle)_{23}\\
&=[\alpha(a_{0}|000\rangle+a_{1}|011\rangle+a_{2}|022\rangle+a_{3}|033\rangle)\\
\nonumber
&~~~~+\beta(a_{0}|100\rangle+a_{1}|111\rangle+a_{2}|122\rangle+a_{3}|133\rangle)\\
\nonumber
&~~~~+\gamma(a_{0}|200\rangle+a_{1}|211\rangle+a_{2}|222\rangle+a_{3}|233\rangle)]_{123}.
\end{align}

Next, we carefully construct the measurement basis of particles 1, 2
\begin{equation}\label{base}
\begin{aligned}
&|\varphi_{1}\rangle_{12}=\frac{1}{\sqrt{3}}(\sqrt{\frac{3}{2}}|00\rangle+\frac{\sqrt{3}}{2}|11\rangle+\frac{\sqrt{3}}{2}|22\rangle)_{12},
&|\varphi_{2}\rangle_{12}=\frac{1}{\sqrt{3}}(\sqrt{\frac{3}{2}}|00\rangle-\frac{\sqrt{3}}{2}|11\rangle-\frac{\sqrt{3}}{2}|22\rangle)_{12},\\
&|\varphi_{3}\rangle_{12}=\frac{1}{\sqrt{3}}(\sqrt{\frac{3}{2}}|01\rangle+\frac{\sqrt{3}}{2}|12\rangle+\frac{\sqrt{3}}{2}|23\rangle)_{12},
&|\varphi_{4}\rangle_{12}=\frac{1}{\sqrt{3}}(\sqrt{\frac{3}{2}}|01\rangle-\frac{\sqrt{3}}{2}|12\rangle-\frac{\sqrt{3}}{2}|23\rangle)_{12},\\
&|\varphi_{5}\rangle_{12}=\frac{1}{\sqrt{3}}(\sqrt{\frac{3}{2}}|02\rangle+\frac{\sqrt{3}}{2}|13\rangle+\frac{\sqrt{3}}{2}|20\rangle)_{12},
&|\varphi_{6}\rangle_{12}=\frac{1}{\sqrt{3}}(\sqrt{\frac{3}{2}}|02\rangle-\frac{\sqrt{3}}{2}|13\rangle-\frac{\sqrt{3}}{2}|20\rangle)_{12},\\
&|\varphi_{7}\rangle_{12}=\frac{1}{\sqrt{3}}(\sqrt{\frac{3}{2}}|03\rangle+\frac{\sqrt{3}}{2}|10\rangle+\frac{\sqrt{3}}{2}|21\rangle)_{12},
&|\varphi_{8}\rangle_{12}=\frac{1}{\sqrt{3}}(\sqrt{\frac{3}{2}}|03\rangle-\frac{\sqrt{3}}{2}|10\rangle-\frac{\sqrt{3}}{2}|21\rangle)_{12},\\
&|\varphi_{9}\rangle_{12}=\frac{1}{\sqrt{2}}(|11\rangle-|22\rangle)_{12},
&|\varphi_{10}\rangle_{12}=\frac{1}{\sqrt{2}}(|12\rangle-|23\rangle)_{12},\\
&|\varphi_{11}\rangle_{12}=\frac{1}{\sqrt{2}}(|13\rangle-|20\rangle)_{12},
&|\varphi_{12}\rangle_{12}=\frac{1}{\sqrt{2}}(|10\rangle-|21\rangle)_{12}, \\
\end{aligned}
\end{equation}
which is vital for the realization of teleportation.

To distinguish the twelve states in Eq. (\ref{base}), we first measure them using a set of projection operators $\{M_{1},M_{2},M_{3},M_{4}\}$, where
\begin{equation}
\begin{aligned}
M_{1}=|00\rangle\langle00|+|11\rangle\langle11|+|22\rangle\langle22|,
&~~~~~M_{2}=|01\rangle\langle01|+|12\rangle\langle12|+|23\rangle\langle23|,\\
M_{3}=|02\rangle\langle02|+|13\rangle\langle13|+|20\rangle\langle20|,
&~~~~~M_{4}=|03\rangle\langle03|+|10\rangle\langle10|+|21\rangle\langle21|.\\
\end{aligned}
\end{equation}
Thus, these twelve quantum states are projected into different subspaces $\{|00\rangle,|11\rangle,|22\rangle\}$, $\{|01\rangle,|12\rangle,|23\rangle\}$, $\{|02\rangle,|13\rangle,|20\rangle\}$, and $\{|03\rangle,|10\rangle,|21\rangle\}$ respectively. Then, the quantum states in different subspaces are measured respectively.

In space $\{|00\rangle, |11\rangle, |22\rangle\}$, measure the quantum states therein using a pair of measurement operators
\begin{equation}
\begin{aligned}
&Q_{1}=\frac{1}{2}(|11\rangle-|22\rangle)(\langle11|-\langle22|), &Q_{2}=I-Q_{1}.
\end{aligned}
\end{equation}
If the measurement result is $Q_{1}$ then the quantum state becomes $|\varphi_{9}\rangle_{12}=\frac{1}{\sqrt{2}}(|11\rangle-|22\rangle)_{12}$; when the measurement result is $Q_{2}$, the quantum states are $|\varphi_{1}\rangle_{12}=\frac{1}{\sqrt{3}}(\sqrt{\frac{3}{2}}|00\rangle+\frac{\sqrt{3}}{2}|11\rangle+\frac{\sqrt{3}}{2}|22\rangle)_{12}$ or $|\varphi_{2}\rangle_{12}=\frac{1}{\sqrt{3}}(\sqrt{\frac{3}{2}}|00\rangle-\frac{\sqrt{3}}{2}|11\rangle-\frac{\sqrt{3}}{2}|22\rangle)_{12}$.
After performing a measurement with the quantum measurement operators
\begin{equation}
\begin{aligned}
&R_{1}=|\varphi_{1}\rangle_{12}\langle\varphi_{1}|,
&R_{2}=|\varphi_{2}\rangle_{12}\langle\varphi_{2}|,
\end{aligned}
\end{equation}
we can know which the final quantum state is obtained. Thus we have completed the state discrimination. The measurements in the other subspaces are similar. By using the basis (\ref{base}) to perform joint measurements on particles 1 and 2 held in Alice's hand, the unnormalized collapsed states of particle 3 corresponding to Alice's joint measurement basis states are $|\phi_{j}\rangle_{3}=_{12}\langle\varphi_{j}|\Psi\rangle_{123}$, which are
\begin{equation}\label{base2}\nonumber
\begin{aligned}
&|\phi_{1}\rangle_{3}=\frac{1}{\sqrt{3}}[\alpha\sqrt{\frac{3}{2}}a_{0}|0\rangle+\beta\frac{\sqrt{3}}{2}a_{1}|1\rangle+\gamma\frac{\sqrt{3}}{2}a_{2}|2\rangle]_{3},\\
&|\phi_{2}\rangle_{3}=\frac{1}{\sqrt{3}}[\alpha\sqrt{\frac{3}{2}}a_{0}|0\rangle-\beta\frac{\sqrt{3}}{2}a_{1}|1\rangle-\gamma\frac{\sqrt{3}}{2}a_{2}|2\rangle]_{3},\\
&|\phi_{3}\rangle_{3}=\frac{1}{\sqrt{3}}[\alpha\sqrt{\frac{3}{2}}a_{1}|1\rangle+\beta\frac{\sqrt{3}}{2}a_{2}|2\rangle+\gamma\frac{\sqrt{3}}{2}a_{3}|3\rangle]_{3},\\
&|\phi_{4}\rangle_{3}=\frac{1}{\sqrt{3}}[\alpha\sqrt{\frac{3}{2}}a_{1}|1\rangle-\beta\frac{\sqrt{3}}{2}a_{2}|2\rangle-\gamma\frac{\sqrt{3}}{2}a_{3}|3\rangle]_{3},\\
\end{aligned}
\end{equation}
\begin{equation}
\begin{aligned}
&|\phi_{5}\rangle_{3}=\frac{1}{\sqrt{3}}[\alpha\sqrt{\frac{3}{2}}a_{2}|2\rangle+\beta\frac{\sqrt{3}}{2}a_{3}|3\rangle+\gamma\frac{\sqrt{3}}{2}a_{0}|0\rangle]_{3},\\
&|\phi_{6}\rangle_{3}=\frac{1}{\sqrt{3}}[\alpha\sqrt{\frac{3}{2}}a_{2}|2\rangle-\beta\frac{\sqrt{3}}{2}a_{3}|3\rangle-\gamma\frac{\sqrt{3}}{2}a_{0}|0\rangle]_{3},\\
&|\phi_{7}\rangle_{3}=\frac{1}{\sqrt{3}}[\alpha\sqrt{\frac{3}{2}}a_{3}|3\rangle+\beta\frac{\sqrt{3}}{2}a_{0}|0\rangle+\gamma\frac{\sqrt{3}}{2}a_{1}|1\rangle]_{3},\\
&|\phi_{8}\rangle_{3}=\frac{1}{\sqrt{3}}[\alpha\sqrt{\frac{3}{2}}a_{3}|3\rangle-\beta\frac{\sqrt{3}}{2}a_{0}|0\rangle-\gamma\frac{\sqrt{3}}{2}a_{1}|1\rangle]_{3},\\
&|\phi_{9}\rangle_{3}=\frac{1}{\sqrt{2}}[\beta a_{1}|1\rangle-\gamma a_{2}|2\rangle]_{3},\\
&|\phi_{10}\rangle_{3}=\frac{1}{\sqrt{2}}[\beta a_{2}|2\rangle-\gamma a_{3}|3\rangle]_{3},\\
&|\phi_{11}\rangle_{3}=\frac{1}{\sqrt{2}}[\beta a_{3}|3\rangle-\gamma a_{0}|0\rangle]_{3},\\
&|\phi_{12}\rangle_{3}=\frac{1}{\sqrt{2}}[\beta a_{0}|0\rangle-\gamma a_{1}|1\rangle]_{3}.\\
\end{aligned}
\end{equation}
The probability $P_{j}$ of Alice's measurement result being $|\varphi_{j}\rangle$, which is also equal to the probability to obtained quantum state $|\phi_{j}\rangle$ is   $P_{j}=_{123}\langle\Psi|\varphi_{j}\rangle\langle\varphi_{j}|\Psi\rangle_{123}$. We can easily to derive
\begin{equation}
\begin{aligned}
&P_{1}=P_{2}=\frac{1}{2}(|\alpha|^{2}a_{0}^{2}+|\beta|^{2}\frac{a_{1}^{2}}{2}+|\gamma|^{2}\frac{a_{2}^{2}}{2}),
&P_{3}=P_{4}=\frac{1}{2}(|\alpha|^{2}a_{1}^{2}+|\beta|^{2}\frac{a_{2}^{2}}{2}+|\gamma|^{2}\frac{a_{3}^{2}}{2}),\\
&P_{5}=P_{6}=\frac{1}{2}(|\alpha|^{2}a_{2}^{2}+|\beta|^{2}\frac{a_{3}^{2}}{2}+|\gamma|^{2}\frac{a_{0}^{2}}{2}),
&P_{7}=P_{8}=\frac{1}{2}(|\alpha|^{2}a_{3}^{2}+|\beta|^{2}\frac{a_{0}^{2}}{2}+|\gamma|^{2}\frac{a_{1}^{2}}{2}),\\
&P_{9}=\frac{1}{2}(|\beta|^{2}a_{1}^{2}+|\gamma|^{2}a_{2}^{2}),
&P_{10}=\frac{1}{2}(|\beta|^{2}a_{2}^{2}+|\gamma|^{2}a_{3}^{2}),\\
&P_{11}=\frac{1}{2}(|\beta|^{2}a_{3}^{2}+|\gamma|^{2}a_{0}^{2}),
&P_{12}=\frac{1}{2}(|\beta|^{2}a_{0}^{2}+|\gamma|^{2}a_{1}^{2}).\\
\end{aligned}
\end{equation}

Obviously, when Alice's measurement outcomes are $|\varphi_{1}\rangle, ... , |\varphi_{8}\rangle$ with the probability $P_{1}, ... , P_{8}$, respectively, the forms of states of particle 3 satisfy condition (ii), i.e.,$|\phi_{j}\rangle_{3}=\alpha|\widetilde{0}\rangle+\beta|\widetilde{1}\rangle+\gamma|\widetilde{2}\rangle$, one can precede the teleportation. On the contrary, if Alice's measurement results are $|\varphi_{9}\rangle, ... , |\varphi_{12}\rangle$ with the probability $P_{9}, ... , P_{12}$, respectively, the teleportation fails. Alice informs Bob of her measurement outcomes  via a classical channel. Then, in order to perfectly transmit quantum state $|\psi\rangle_{1}$, Bob will perform appropriate local operations related to Alice's measurement outcomes. The detailed procedure will be stated in the next section.

\begin{figure}[h]
\centering
\includegraphics[width=0.55\textwidth]{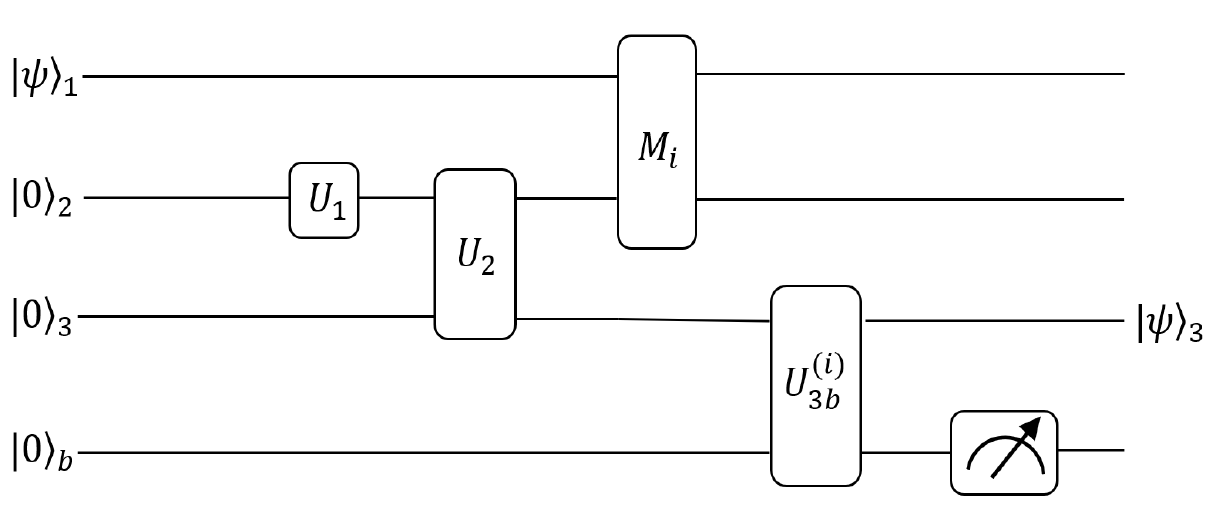}
\caption{The detailed quantum circuit of the quantum teleportation scheme is as follows. Initially, a qutrit (particle 1) is in an unknown quantum state $|\psi\rangle_1$ to be teleported. Both 4-level particles 2 and 3 are in the state $|0\rangle$. By performing a unitary operator $U_{1}$ on Alice's particle 2 and a joint unitary operator $U_2$ on particles 2 and 3, and sending the particle 3 to the receiver Bob, the quantum channel has been prepared. Then Alice makes a quantum measurement $\{M_i\}$ on the particles 1 and 2, and informs the receiver Bob her measurement outcome via a classical information channel. According to Alice's measurement result, Bob operates the corresponding collective unitary operator $U_{3b}^{(i)}$ on particle 3 and an auxiliary qubit $b$ with initial state $|0\rangle_{b}$. After that Bob makes a measurement on the auxiliary qubit $b$. If the measurement outcome is $|0\rangle_b$, then the quantum state of particle 3 is the unknown quantum state $|\psi\rangle$. The teleportation has been successfully realized. \label{fig1}}
\end{figure}

The quantum circuit shown in Fig.\ref{fig1} provides a more detailed description of the quantum teleportation.

\section{Reconstruct the unkonwn state to be transmitted}\label{Q3}

According to Alice's different measurement results, Bob obtains the corresponding state of particle 3, and then performs the corresponding local operations to reconstruct the unknown state. In order to fully realize quantum teleportation, Bob needs to introduce an auxiliary qubit $b$ with initial state $|0\rangle_{b}$ and performs a collective unitary transformation. For example, when Alice's measurement result is
$|\varphi_{1}\rangle_{12}=\frac{1}{\sqrt{3}}(\sqrt{\frac{3}{2}}|00\rangle+\frac{\sqrt{3}}{2}|11\rangle+\frac{\sqrt{3}}{2}|22\rangle)_{12}$, particle 3 is in state
$|\phi_{1}\rangle_{3}=\frac{[\alpha\sqrt{\frac{3}{2}}a_{0}|0\rangle+\beta\frac{\sqrt{3}}{2}a_{1}|1\rangle+\gamma\frac{\sqrt{3}}{2}a_{2}|2\rangle]_{3}}{\sqrt{3(2|\alpha|^{2}a_{0}^{^{2}}+|\beta|^{2}a_{1}^{2}+|\gamma|^{2}a_{2}^{2})}}$. In the basis $\{|00\rangle_{3b}, |01\rangle_{3b}, |10\rangle_{3b}, |11\rangle_{3b}, |20\rangle_{3b}, |21\rangle_{3b}\}$, the quantum state $|\phi_{1}\rangle_{3}|0\rangle_b$ can be expressed as
\begin{equation}\label{base3}
\begin{aligned}
\frac{[\alpha\sqrt{\frac{3}{2}}a_{0}|0\rangle+\beta\frac{\sqrt{3}}{2}a_{1}|1\rangle+\gamma\frac{\sqrt{3}}{2}a_{2}|2\rangle]_{3}}{\sqrt{3(2|\alpha|^{2}a_{0}^{^{2}}+|\beta|^{2}a_{1}^{2}+|\gamma|^{2}a_{2}^{2})}}|0\rangle_{b}=
\left(\begin{array}{c}
\frac{\alpha\sqrt{2}a_{0}}{\sqrt{2|\alpha|^{2}a_{0}^{^{2}}+|\beta|^{2}a_{1}^{2}+|\gamma|^{2}a_{2}^{2}}}\\
        0\\
      \frac{\beta a_{1}}
      {\sqrt{2|\alpha|^{2}a_{0}^{^{2}}+|\beta|^{2}a_{1}^{2}+|\gamma|^{2}a_{2}^{2}}}\\
        0\\
        \frac{\gamma a_{2}}
        {\sqrt{2|\alpha|^{2}a_{0}^{^{2}}+|\beta|^{2}a_{1}^{2}+|\gamma|^{2}a_{2}^{2}}}\\
        0\\
\end{array}
\right).
\end{aligned}
\end{equation}
If $\sqrt{2}a_{0}<a_{1}$, Bob performs a joint unitary operation
\begin{equation}
\begin{aligned}
U_{3b}^{(1)} =\left(\begin{array}{cccccc}
        1 & 0 & 0 & 0 & 0 & 0 \\
        0 & 1 & 0 & 0 & 0 & 0 \\
        0 & 0 & \frac{\sqrt{2}a_{0}}{a_{1}} & \sqrt{1-\frac{2a_{0}^{2}}{a_{1}^{2}}} & 0 & 0 \\
        0 & 0 & -\sqrt{1-\frac{2a_{0}^{2}}{a_{1}^{2}}} & \frac{\sqrt{2}a_{0}}{a_{1}} & 0 & 0 \\
        0 & 0 & 0 & 0 & \frac{\sqrt{2}a_{0}}{a_{2}} & \sqrt{1-\frac{2a_{0}^{2}}{a_{2}^{2}}} \\
        0 & 0 & 0 & 0 & -\sqrt{1-\frac{2a_{0}^{2}}{a_{2}^{2}}} & \frac{\sqrt{2}a_{0}}{a_{2}} \\
\end{array}
\right)
\end{aligned}
\end{equation}
on the particles 3 and $b$. Thus we have
\begin{equation}
\begin{aligned}
&U_{3b}^{(1)}
    \begin{pmatrix}
        \frac{\alpha\sqrt{2}a_{0}}{\sqrt{2|\alpha|^{2}a_{0}^{^{2}}+|\beta|^{2}a_{1}^{2}+|\gamma|^{2}a_{2}^{2}}}\\
        0\\
               \frac{\beta a_{1}}{\sqrt{2|\alpha|^{2}a_{0}^{^{2}}+|\beta|^{2}a_{1}^{2}+|\gamma|^{2}a_{2}^{2}}}\\
        0\\
                \frac{\gamma a_{2}}{\sqrt{2|\alpha|^{2}a_{0}^{^{2}}+|\beta|^{2}a_{1}^{2}+|\gamma|^{2}a_{2}^{2}}}\\
        0\\
    \end{pmatrix}
       &=\frac{\sqrt{2}a_{0}(\alpha|0\rangle+\beta|1\rangle+\gamma|2\rangle)_{3}|0\rangle_{b}+
    (-\beta\sqrt{a_{1}^{2}-2a_{0}^{2}}|1\rangle-\gamma\sqrt{a_{2}^{2}-2a_{0}^{2}}|2\rangle)_{3}|1\rangle_{b}}{\sqrt{2|\alpha|^{2}a_{0}^{^{2}}+|\beta|^{2}a_{1}^{2}+|\gamma|^{2}a_{2}^{2}}}.
\end{aligned}
\end{equation}
If $\sqrt{2}a_{0}>a_{1}$, then we use
\begin{equation}
\begin{aligned}
U_{3b}^{'(1)} =
    \begin{pmatrix}
        \frac{a_{1}}{\sqrt{2}a_{0}} & \sqrt{1-\frac{a_{1}^{2}}{2a_{0}^{2}}} & 0 & 0 & 0 & 0 \\
        -\sqrt{1-\frac{a_{1}^{2}}{2a_{0}^{2}}} & \frac{a_{1}}{\sqrt{2}a_{0}} & 0 & 0 & 0 & 0 \\
        0 & 0 & 1 & 0 & 0 & 0 \\
        0 & 0 & 0 & 1 & 0 & 0 \\
        0 & 0 & 0 & 0 & \frac{a_{1}}{a_{2}} & \sqrt{1-\frac{a_{1}^{2}}{a_{2}^{2}}} \\
        0 & 0 & 0 & 0 & -\sqrt{1-\frac{a_{1}^{2}}{a_{2}^{2}}} & \frac{a_{1}}{a_{2}} \\
    \end{pmatrix}
\end{aligned}
\end{equation}
\begin{table}[tbp]
\centering
\caption{ Alice's different measurement results, the states of particle 3, the collective unitary transformations performed by Bob on auxiliary particles $b$ and  3, as well as the states of particles $b$ and 3 after the transformation are performed.}
\label{T1}

\begin{tabular}{c c c c}
\hline
\hline

\tabincell{c}{Alice's measurement\\result}
&
~~~~~~~~State of particle 3~~~~~
&
~~~~~~\tabincell{c}{The collective unitary\\ transformation performed by Bob}~~~~~~
&
\tabincell{c}{State of particles $b$ and 3 after Bob's joint\\unitary transformation is performed}
\\

\hline

$|\varphi_{1}\rangle_{12}$ & $|\phi_{1}\rangle_{3}$ & $U_{3b}^{(1)}/U_{3b}^{'(1)}$ & $|\chi_{1}\rangle/|\chi_{1}^{'}\rangle$ \\
$|\varphi_{2}\rangle_{12}$ & $|\phi_{2}\rangle_{3}$ & $U_{3b}^{(2)}/U_{3b}^{'(2)}$ & $|\chi_{2}\rangle/|\chi_{2}^{'}\rangle$ \\
$|\varphi_{3}\rangle_{12}$ & $|\phi_{3}\rangle_{3}$ & $U_{3b}^{(3)}/U_{3b}^{'(3)}$ & $|\chi_{3}\rangle/|\chi_{3}^{'}\rangle$ \\
$|\varphi_{4}\rangle_{12}$ & $|\phi_{4}\rangle_{3}$ & $U_{3b}^{(4)}/U_{3b}^{'(4)}$ & $|\chi_{4}\rangle/|\chi_{4}^{'}\rangle$ \\
$|\varphi_{5}\rangle_{12}$ & $|\phi_{5}\rangle_{3}$ & $U_{3b}^{(5)}$ & $|\chi_{5}\rangle$ \\
$|\varphi_{6}\rangle_{12}$ & $|\phi_{6}\rangle_{3}$ & $U_{3b}^{(6)}$ & $|\chi_{6}\rangle$ \\
$|\varphi_{7}\rangle_{12}$ & $|\phi_{7}\rangle_{3}$ & $U_{3b}^{(7)}$ & $|\chi_{7}\rangle$ \\
$|\varphi_{8}\rangle_{12}$ & $|\phi_{8}\rangle_{3}$ & $U_{3b}^{(8)}$ & $|\chi_{8}\rangle$ \\

\hline
\hline

\end{tabular}
\end{table}
acting on quantum state (\ref{base3}) to give
\begin{equation}
\begin{aligned}
&U_{3b}^{'(1)}
    \begin{pmatrix}
        \frac{\alpha\sqrt{2}a_{0}}{\sqrt{2|\alpha|^{2}a_{0}^{^{2}}+|\beta|^{2}a_{1}^{2}+|\gamma|^{2}a_{2}^{2}}}\\
        0\\
        \frac{\beta a_{1}}{\sqrt{2|\alpha|^{2}a_{0}^{^{2}}+|\beta|^{2}a_{1}^{2}+|\gamma|^{2}a_{2}^{2}}}\\
        0\\
        \frac{\gamma a_{2}}{\sqrt{2|\alpha|^{2}a_{0}^{^{2}}+|\beta|^{2}a_{1}^{2}+|\gamma|^{2}a_{2}^{2}}}\\
        0\\
        \end{pmatrix}
        &=\frac{a_{1}(\alpha|0\rangle+\beta|1\rangle+\gamma|2\rangle)_{3}|0\rangle_{b}+
    (-\alpha\sqrt{2a_{0}^{2}-a_{1}^{2}}|0\rangle-\gamma\sqrt{a_{2}^{2}-a_{1}^{2}}|2\rangle)_{3}|1\rangle_b}{\sqrt{2|\alpha|^{2}a_{0}^{^{2}}+|\beta|^{2}a_{1}^{2}+|\gamma|^{2}a_{2}^{2}}}.
\end{aligned}
\end{equation}
Next, one performs a von Neumann measurement on the introduced auxiliary particle $b$. If the measurement result of auxiliary particle $b$ is $|0\rangle_{b}$, the teleportation is successful; if the measurement result is $|1\rangle_{b}$, then the teleportation fails.

If Alice's measurement result is one of the other cases, then the particle 3 in Bob's hand will collapse into a corresponding state in Eq. (\ref{base2}). We can subsequently use the similar method to achieve the final goal. Table \ref{T1} presents the state of particle 3, the collective unitary transformation applied by Bob to auxiliary particle $b$ and particle 3, and the post-transformation states of particles $b$ and 3 for Alice's different measurement results.

In Table \ref{T1}, when $\sqrt{2}a_{0}<a_{1}$,
\begin{equation}
\begin{aligned}
U_{3b}^{(1)} =
    \begin{pmatrix}
        1 & 0 & 0 & 0 & 0 & 0 \\
        0 & 1 & 0 & 0 & 0 & 0 \\
        0 & 0 & \frac{2a_{0}}{\sqrt{2}a_{1}} & \sqrt{1-\frac{2a_{0}^{2}}{a_{1}^{2}}} & 0 & 0 \\
        0 & 0 & -\sqrt{1-\frac{2a_{0}^{2}}{a_{1}^{2}}} & \frac{2a_{0}}{\sqrt{2}a_{1}} & 0 & 0 \\
        0 & 0 & 0 & 0 & \frac{2a_{0}}{\sqrt{2}a_{2}} & \sqrt{1-\frac{2a_{0}^{2}}{a_{2}^{2}}} \\
        0 & 0 & 0 & 0 & -\sqrt{1-\frac{2a_{0}^{2}}{a_{2}^{2}}} & \frac{2a_{0}}{\sqrt{2}a_{2}} \\
    \end{pmatrix}\\
\end{aligned},
\end{equation}
when $\sqrt{2}a_{0}>a_{1}$,
\begin{equation}
\begin{aligned}
U_{3b}^{'(1)} =
    \begin{pmatrix}
        \frac{a_{1}}{\sqrt{2}a_{0}} & \sqrt{1-\frac{a_{1}^{2}}{2a_{0}^{2}}} & 0 & 0 & 0 & 0 \\
        -\sqrt{1-\frac{a_{1}^{2}}{2a_{0}^{2}}} & \frac{a_{1}}{\sqrt{2}a_{0}} & 0 & 0 & 0 & 0 \\
        0 & 0 & 1 & 0 & 0 & 0 \\
        0 & 0 & 0 & 1 & 0 & 0 \\
        0 & 0 & 0 & 0 & \frac{a_{1}}{a_{2}} & \sqrt{1-\frac{a_{1}^{2}}{a_{2}^{2}}} \\
        0 & 0 & 0 & 0 & -\sqrt{1-\frac{a_{1}^{2}}{a_{2}^{2}}} & \frac{a_{1}}{a_{2}} \\
    \end{pmatrix};
\end{aligned}
\end{equation}
when $\sqrt{2}a_{0}<a_{1}$,
\begin{equation}
\begin{aligned}
U_{3b}^{(2)} =
    \begin{pmatrix}
        1 & 0 & 0 & 0 & 0 & 0 \\
        0 & 1 & 0 & 0 & 0 & 0 \\
        0 & 0 & -\frac{\sqrt{2}a_{0}}{a_{1}} & \sqrt{1-\frac{2a_{0}^{2}}{a_{1}^{2}}} & 0 & 0 \\
        0 & 0 & -\sqrt{1-\frac{2a_{0}^{2}}{a_{1}^{2}}} & -\frac{\sqrt{2}a_{0}}{a_{1}} & 0 & 0 \\
        0 & 0 & 0 & 0 & -\frac{\sqrt{2}a_{0}}{a_{2}} & \sqrt{1-\frac{2a_{0}^{2}}{a_{2}^{2}}} \\
        0 & 0 & 0 & 0 & -\sqrt{1-\frac{2a_{0}^{2}}{a_{2}^{2}}} & -\frac{\sqrt{2}a_{0}}{a_{2}} \\
    \end{pmatrix},\\
\end{aligned}
\end{equation}
when $\sqrt{2}a_{0}>a_{1}$,
\begin{equation}
\begin{aligned}
U_{3b}^{'(2)} =
    \begin{pmatrix}
        \frac{a_{1}}{\sqrt{2}a_{0}} & \sqrt{1-\frac{a_{1}^{2}}{2a_{0}^{2}}} & 0 & 0 & 0 & 0 \\
        -\sqrt{1-\frac{a_{1}^{2}}{2a_{0}^{2}}} & \frac{a_{1}}{\sqrt{2}a_{0}} & 0 & 0 & 0 & 0 \\
        0 & 0 & -1 & 0 & 0 & 0 \\
        0 & 0 & 0 & 1 & 0 & 0 \\
        0 & 0 & 0 & 0 & -\frac{a_{1}}{a_{2}} & \sqrt{1-\frac{a_{1}^{2}}{a_{2}^{2}}} \\
        0 & 0 & 0 & 0 & -\sqrt{1-\frac{a_{1}^{2}}{a_{2}^{2}}} & -\frac{a_{1}}{a_{2}} \\
    \end{pmatrix};
\end{aligned}
\end{equation}
when $\sqrt{2}a_{1}<a_{2}$,
\begin{equation}
\begin{aligned}
U_{3b}^{(3)} =
    \begin{pmatrix}
        1 & 0 & 0 & 0 & 0 & 0\\
        0 & 1 & 0 & 0 & 0 & 0\\
        0 & 0 & \frac{2a_{1}}{\sqrt{2}a_{2}} & \sqrt{1-\frac{2a_{1}^{2}}{a_{2}^{2}}} & 0 & 0 \\
        0 & 0 & -\sqrt{1-\frac{2a_{1}^{2}}{a_{2}^{2}}} & \frac{2a_{1}}{\sqrt{2}a_{2}} & 0 & 0 \\
        0 & 0 & 0 & 0 & \frac{2a_{1}}{\sqrt{2}a_{3}} & \sqrt{1-\frac{2a_{1}^{2}}{a_{3}^{2}}} \\
        0 & 0 & 0 & 0 & -\sqrt{1-\frac{2a_{1}^{2}}{a_{3}^{2}}} & \frac{2a_{1}}{\sqrt{2}a_{3}} \\
    \end{pmatrix},\\
\end{aligned}
\end{equation}
when $\sqrt{2}a_{1}>a_{2}$,
\begin{equation}
\begin{aligned}
U_{3b}^{'(3)} =
    \begin{pmatrix}
        \frac{a_{2}}{\sqrt{2}a_{1}} & \sqrt{1-\frac{a_{2}^{2}}{2a_{1}^{2}}} & 0 & 0 & 0 & 0\\
        -\sqrt{1-\frac{a_{2}^{2}}{2a_{1}^{2}}} & \frac{a_{2}}{\sqrt{2}a_{1}} & 0 & 0 & 0 & 0\\
        0 & 0 & 1 & 0 & 0 & 0 \\
        0 & 0 & 0 & 1 & 0 & 0 \\
        0 & 0 & 0 & 0 & \frac{a_{2}}{a_{3}} & \sqrt{1-\frac{a_{2}^{2}}{a_{3}^{2}}}\\
        0 & 0 & 0 & 0 & -\sqrt{1-\frac{a_{2}^{2}}{a_{3}^{2}}} & \frac{a_{2}}{a_{3}}\\
    \end{pmatrix};\\
\end{aligned}
\end{equation}
when $\sqrt{2}a_{1}<a_{2}$,
\begin{equation}
\begin{aligned}
U_{3b}^{(4)} =
    \begin{pmatrix}
        1 & 0 & 0 & 0 & 0 & 0\\
        0 & 1 & 0 & 0 & 0 & 0\\
        0 & 0 & -\frac{2a_{1}}{\sqrt{2}a_{2}} & \sqrt{1-\frac{2a_{1}^{2}}{a_{2}^{2}}} & 0 & 0 \\
        0 & 0 & -\sqrt{1-\frac{2a_{1}^{2}}{a_{2}^{2}}} & \frac{2a_{1}}{\sqrt{2}a_{2}} & 0 & 0 \\
        0 & 0 & 0 & 0 & -\frac{2a_{1}}{\sqrt{2}a_{3}} & \sqrt{1-\frac{2a_{1}^{2}}{a_{3}^{2}}} \\
        0 & 0 & 0 & 0 & -\sqrt{1-\frac{2a_{1}^{2}}{a_{3}^{2}}} & \frac{2a_{1}}{\sqrt{2}a_{3}} \\
    \end{pmatrix},\\
\end{aligned}
\end{equation}
when $\sqrt{2}a_{1}>a_{2}$,
\begin{equation}
\begin{aligned}
U_{3b}^{'(4)} =
    \begin{pmatrix}
        \frac{a_{2}}{\sqrt{2}a_{1}} & \sqrt{1-\frac{a_{2}^{2}}{2a_{1}^{2}}} & 0 & 0 & 0 & 0\\
        -\sqrt{1-\frac{a_{2}^{2}}{2a_{1}^{2}}} & \frac{a_{2}}{\sqrt{2}a_{1}} & 0 & 0 & 0 & 0\\
        0 & 0 & -1 & 0 & 0 & 0\\
        0 & 0 & 0 & 0 & 0 & 0 \\
        0 & 0 & 0 & 0 & -\frac{a_{2}}{a_{3}} & \sqrt{1-\frac{a_{2}^{2}}{a_{3}^{2}}}\\
        0 & 0 & 0 & 0 & -\sqrt{1-\frac{a_{2}^{2}}{a_{3}^{2}}} & -\frac{a_{2}}{a_{3}}\\
    \end{pmatrix};\\
\end{aligned}
\end{equation}

\begin{equation}
\begin{aligned}
U_{3b}^{(5)} =
    \begin{pmatrix}
        0 & 0 & \frac{a_{0}}{\sqrt{2}a_{2}} & \sqrt{1-\frac{a_{0}^{2}}{2a_{2}^{2}}} & 0 & 0\\
        0 & 0 & -\sqrt{1-\frac{a_{0}^{2}}{2a_{2}^{2}}} & \frac{a_{0}}{\sqrt{2}a_{2}} & 0 & 0\\
        0 & 0 & 0 & 0 & \frac{a_{0}}{a_{3}} & \sqrt{1-\frac{a_{0}^{2}}{a_{3}^{2}}}\\
        0 & 0 & 0 & 0 & -\sqrt{1-\frac{a_{0}^{2}}{a_{3}^{2}}} & \frac{a_{0}}{a_{3}}\\
        1 & 0 & 0 & 0 & 0 & 0 \\
        0 & 1 & 0 & 0 & 0 & 0 \\
    \end{pmatrix},\\
\end{aligned}
\end{equation}

\begin{equation}
\begin{aligned}
U_{3b}^{(6)} =
    \begin{pmatrix}
        0 & 0 & \frac{a_{0}}{\sqrt{2}a_{2}} & \sqrt{1-\frac{a_{0}^{2}}{2a_{2}^{2}}} & 0 & 0\\
        0 & 0 & -\sqrt{1-\frac{a_{0}^{2}}{2a_{2}^{2}}} & \frac{a_{0}}{\sqrt{2}a_{2}} & 0 & 0\\
        0 & 0 & 0 & 0 & -\frac{a_{0}}{a_{3}} & \sqrt{1-\frac{a_{0}^{2}}{a_{3}^{2}}}\\
        0 & 0 & 0 & 0 & -\sqrt{1-\frac{a_{0}^{2}}{a_{3}^{2}}} & -\frac{a_{0}}{a_{3}}\\
        1 & 0 & 0 & 0 & 0 & 0\\
        0 & 1 & 0 & 0 & 0 & 0\\
    \end{pmatrix},\\
\end{aligned}
\end{equation}

\begin{equation}
\begin{aligned}
U_{3b}^{(7)} =
    \begin{pmatrix}
        0 & 0 & 0 & 0 & \frac{a_{0}}{\sqrt{2}a_{3}} & \sqrt{1-\frac{a_{0}^{2}}{2a_{3}^{2}}}\\
        0 & 0 & 0 & 0 & -\sqrt{1-\frac{a_{0}^{2}}{2a_{3}^{2}}} & \frac{a_{0}}{\sqrt{2}a_{3}}\\
        1 & 0 & 0 & 0 & 0 & 0\\
        0 & 1 & 0 & 0 & 0 & 0 \\
        0 & 0 & \frac{a_{0}}{a_{1}} & \sqrt{1-\frac{a_{0}^{2}}{a_{1}^{2}}} & 0 & 0\\
        0 & 0 & -\sqrt{1-\frac{a_{0}^{2}}{a_{1}^{2}}} & \frac{a_{0}}{a_{1}} & 0 & 0\\
    \end{pmatrix},\\
\end{aligned}
\end{equation}

\begin{equation}
\begin{aligned}
U_{3b}^{(8)} =
    \begin{pmatrix}
        0 & 0 & 0 & 0 & \frac{a_{0}}{\sqrt{2}a_{3}} & \sqrt{1-\frac{a_{0}^{2}}{2a_{3}^{2}}}\\
        0 & 0 & 0 & 0 & -\sqrt{1-\frac{a_{0}^{2}}{2a_{3}^{2}}} & \frac{a_{0}}{\sqrt{2}a_{3}}\\
        -1 & 0 & 0 & 0 & 0 & 0\\
        0 & 1 & 0 & 0 & 0 & 0\\
        0 & 0 & -\frac{a_{0}}{a_{1}} & \sqrt{1-\frac{a_{0}^{2}}{a_{1}^{2}}} & 0 & 0\\
        0 & 0 & -\sqrt{1-\frac{a_{0}^{2}}{a_{1}^{2}}} & -\frac{a_{0}}{a_{1}} & 0 & 0\\
    \end{pmatrix}.\\
\end{aligned}
\end{equation}

While $|\chi_{1}\rangle,~|\chi'_{1}\rangle,~...,~|\chi_{8}\rangle$ are
\begin{equation}\nonumber
\begin{aligned}
&|\chi_{1}\rangle=\frac{\sqrt{2}a_{0}(\alpha|0\rangle+\beta|1\rangle+\gamma|2\rangle)_{3}|0\rangle_{b}+
    (-\beta\sqrt{a_{1}^{2}-2a_{0}^{2}}|1\rangle-\gamma\sqrt{a_{2}^{2}-2a_{0}^{2}}|2\rangle)_{3}|1\rangle_{b}}{\sqrt{2|\alpha|^{2}a_{0}^{^{2}}+|\beta|^{2}a_{1}^{2}+|\gamma|^{2}a_{2}^{2}}},\\
&|\chi_{1}^{'}\rangle=\frac{a_{1}(\alpha|0\rangle+\beta|1\rangle+\gamma|2\rangle)_{3}|0\rangle_{b}+
    (-\alpha\sqrt{2a_{0}^{2}-a_{1}^{2}}|0\rangle-\gamma\sqrt{a_{2}^{2}-a_{1}^{2}}|2\rangle)_{3}|1\rangle_{b}}{\sqrt{2|\alpha|^{2}a_{0}^{^{2}}+|\beta|^{2}a_{1}^{2}+|\gamma|^{2}a_{2}^{2}}},\\
    \end{aligned}
\end{equation}
\begin{equation}
\begin{aligned}
&|\chi_{2}\rangle=\frac{\sqrt{2}a_{0}(\alpha|0\rangle+\beta|1\rangle+\gamma|2\rangle)_{3}|0\rangle_{b}+
    (\beta\sqrt{a_{1}^{2}-2a_{0}^{2}}|1\rangle+\gamma\sqrt{a_{2}^{2}-2a_{0}^{2}}|2\rangle)_{3}|1\rangle_{b}}{{\sqrt{2|\alpha|^{2}a_{0}^{^{2}}+|\beta|^{2}a_{1}^{2}+|\gamma|^{2}a_{2}^{2}}}},\\
&|\chi_{2}^{'}\rangle=\frac{a_{1}(\alpha|0\rangle+\beta|1\rangle+\gamma|2\rangle)_{3}|0\rangle_{b}+
    (-\alpha\sqrt{2a_{0}^{2}-a_{1}^{2}}|0\rangle+\gamma\sqrt{a_{2}^{2}-a_{1}^{2}}|2\rangle)_{3}|1\rangle_{b}}{{\sqrt{2|\alpha|^{2}a_{0}^{^{2}}+|\beta|^{2}a_{1}^{2}+|\gamma|^{2}a_{2}^{2}}}},\\
&|\chi_{3}\rangle=\frac{\sqrt{2}a_{1}(\alpha|1\rangle+\beta|2\rangle+\gamma|3\rangle)_{3}|0\rangle_{b}+
    (-\beta\sqrt{a_{2}^{2}-2a_{1}^{2}}|2\rangle-\gamma\sqrt{a_{3}^{2}-2a_{1}^{2}}|3\rangle)_{3}|1\rangle_{b}}{{\sqrt{2|\alpha|^{2}a_{1}^{^{2}}+|\beta|^{2}a_{2}^{2}+|\gamma|^{2}a_{3}^{2}}}},\\
&|\chi_{3}^{'}\rangle=\frac{a_{2}(\alpha|1\rangle+\beta|2\rangle+\gamma|3\rangle)_{3}|0\rangle_{b}+
    (-\alpha\sqrt{2a_{1}^{2}-a_{2}^{2}}|1\rangle-\gamma\sqrt{a_{3}^{2}-a_{2}^{2}}|3\rangle)_{3}|1\rangle_{b}}{\sqrt{2|\alpha|^{2}a_{1}^{^{2}}+|\beta|^{2}a_{2}^{2}+|\gamma|^{2}a_{3}^{2}}},\\
&|\chi_{4}\rangle=\frac{\sqrt{2}a_{1}(\alpha|1\rangle+\beta|2\rangle+\gamma|3\rangle)_{3}|0\rangle_{b}+
    (\beta\sqrt{a_{2}^{2}-2a_{1}^{2}}|2\rangle+\gamma\sqrt{a_{3}^{2}-2a_{1}^{2}}|3\rangle)_{3}|1\rangle_{b}}{\sqrt{2|\alpha|^{2}a_{1}^{^{2}}+|\beta|^{2}a_{2}^{2}+|\gamma|^{2}a_{3}^{2}}},\\
&|\chi_{4}^{'}\rangle=\frac{a_{2}(\alpha|1\rangle+\beta|2\rangle+\gamma|3\rangle)_{3}|0\rangle_{b}+
    (-\alpha\sqrt{2a_{1}^{2}-a_{2}^{2}}|1\rangle+\gamma\sqrt{a_{3}^{2}-a_{2}^{2}}|3\rangle)_{3}|1\rangle_{b}}{\sqrt{2|\alpha|^{2}a_{1}^{^{2}}+|\beta|^{2}a_{2}^{2}+|\gamma|^{2}a_{3}^{2}}},\\
&|\chi_{5}\rangle=\frac{a_{0}(\alpha|0\rangle+\beta|2\rangle+\gamma|3\rangle)_{3}|0\rangle_{b}+
    (-\alpha\sqrt{2a_{2}^{2}-a_{0}^{2}}|0\rangle-\beta\sqrt{a_{3}^{2}-a_{0}^{2}}|2\rangle)_{3}|1\rangle_{b}}{\sqrt{2|\alpha|^{2}a_{2}^{^{2}}+|\beta|^{2}a_{3}^{2}+|\gamma|^{2}a_{0}^{2}}},\\
&|\chi_{6}\rangle=\frac{a_{0}(\alpha|0\rangle+\beta|2\rangle+\gamma|3\rangle)_{3}|0\rangle_{b}+
    (-\alpha\sqrt{2a_{2}^{2}-a_{0}^{2}}|0\rangle+\beta\sqrt{a_{3}^{2}-a_{0}^{2}}|2\rangle)_{3}|1\rangle_{b}}{\sqrt{2|\alpha|^{2}a_{2}^{^{2}}+|\beta|^{2}a_{3}^{2}+|\gamma|^{2}a_{0}^{2}}},\\
&|\chi_{7}\rangle=\frac{a_{0}(\alpha|0\rangle+\beta|1\rangle+\gamma|3\rangle)_{3}|0\rangle_{b}+
    (-\alpha\sqrt{2a_{3}^{2}-a_{0}^{2}}|0\rangle-\gamma\sqrt{a_{1}^{2}-a_{0}^{2}}|3\rangle)_{3}|1\rangle_{b}}{\sqrt{2|\alpha|^{2}a_{3}^{^{2}}+|\beta|^{2}a_{0}^{2}+|\gamma|^{2}a_{1}^{2}}},\\
&|\chi_{8}\rangle=\frac{a_{0}(\alpha|0\rangle+\beta|1\rangle+\gamma|3\rangle)_{3}|0\rangle_{b}+
    (-\alpha\sqrt{2a_{3}^{2}-a_{0}^{2}}|0\rangle+\gamma\sqrt{a_{1}^{2}-a_{0}^{2}}|3\rangle)_{3}|1\rangle_{b}}{\sqrt{2|\alpha|^{2}a_{3}^{^{2}}+|\beta|^{2}a_{0}^{2}+|\gamma|^{2}a_{1}^{2}}}.\\
\end{aligned}
\end{equation}

Under the above circumstances, when the quantum state $|\chi_{j}\rangle$ is obtained, the probabilities of successful teleportation, $P_{|\chi_{j}\rangle}$ is respectively as follows
\begin{equation}
\begin{aligned}
&P_{|\chi_{1}\rangle}=P_{|\chi_{2}\rangle}=\frac{2a_{0}^{2}}{2|\alpha|^{2}a_{0}^{^{2}}+|\beta|^{2}a_{1}^{2}+|\gamma|^{2}a_{2}^{2}},
&P_{|\chi_{1}^{'}\rangle}=P_{|\chi_{2}^{'}\rangle}=\frac{a_{1}^{2}}{2|\alpha|^{2}a_{0}^{^{2}}+|\beta|^{2}a_{1}^{2}+|\gamma|^{2}a_{2}^{2}},\\
&P_{|\chi_{3}\rangle}=P_{|\chi_{4}\rangle}=\frac{2a_{1}^{2}}{{2|\alpha|^{2}a_{1}^{^{2}}+|\beta|^{2}a_{2}^{2}+|\gamma|^{2}a_{3}^{2}}},
&P_{|\chi_{3}^{'}\rangle}=P_{|\chi_{4}^{'}\rangle}=\frac{a_{2}^{2}}{{2|\alpha|^{2}a_{1}^{^{2}}+|\beta|^{2}a_{2}^{2}+|\gamma|^{2}a_{3}^{2}}},\\
&P_{|\chi_{5}\rangle}=P_{|\chi_{6}\rangle}=\frac{a_{0}^{2}}{2|\alpha|^{2}a_{2}^{^{2}}+|\beta|^{2}a_{3}^{2}+|\gamma|^{2}a_{0}^{2}},
&P_{|\chi_{7}\rangle}=P_{|\chi_{8}\rangle}=\frac{a_{0}^{2}}{2|\alpha|^{2}a_{3}^{^{2}}+|\beta|^{2}a_{0}^{2}+|\gamma|^{2}a_{1}^{2}}.\\
\end{aligned}
\end{equation}

Next we will investigate the total success probability of the teleportation. Clearly,
when $\sqrt{2}a_{0}<a_{1}$, $\sqrt{2}a_{1}<a_{2}$, the total success probability is
\begin{equation}
P_{total}^{(1)}=2a_{0}^{2}+a_{1}^{2}.
\end{equation}
Similarly, if $\sqrt{2}a_{0}>a_{1}$, $\sqrt{2}a_{1}<a_{2}$, then the total success probability
\begin{equation}
P_{total}^{(2)}=a_{0}^{2}+\frac{3a_{1}^{2}}{2}.
\end{equation}
For the case $\sqrt{2}a_{0}<a_{1}$, $\sqrt{2}a_{1}>a_{2}$, we obtain the total success probability
\begin{equation}
P_{total}^{(3)}=2a_{0}^{2}+\frac{a_{2}^{2}}{2}.
\end{equation}
Under the condition $\sqrt{2}a_{0}>a_{1}$, $\sqrt{2}a_{1}>a_{2}$, the total success probability becomes
\begin{equation}
P_{total}^{(4)}=a_{0}^{2}+\frac{a_{1}^{2}}{2}+\frac{a_{2}^{2}}{2}.
\end{equation}

\section{Discussion and summary}\label{Q4}

In summary, we propose a quantum teleportation scheme to transmit a single qutrit by adopting a 2-qudit quantum entangled state as the quantum channel. We systematically construct Alice's measurement operations for implementing the teleportation protocol. The sender Alice performs the joint measurement on the qutrit and one of the entangled particles, and announces her measurement outcomes. We carefully design the corresponding unitary transformation on another entangled particle and an auxiliary qubit. After the corresponding unitary operation has been implemented and a von Neumann measurement on the auxiliary particle has been carried out, the unknown quantum state of a single qutrit has been perfectly reconstructed with fidelity of 1 and a finite success probability.

In this work, although we establish a teleportation scheme for a qutrit system, the protocol can also be used to transmit a unknown qubit state with a higher success probability. By Eq. (\ref{base2}), we can easily see that when $\alpha=0$, the teleportation protocol reduces to the teleportation for a qubit state. In this situation, all of Alice's measurement results can be used to teleport the unknown quantum state $\beta|1\rangle +\gamma|2\rangle$, which consequently increases the success probability of teleportation.

The obtained results enrich the theory of quantum teleportation over high-dimensional entangled channels and provide a novel implementation method for qutrit teleportation. We expect that this quantum teleportation protocol can be verified by reliable experiments in the future.

\end{document}